\documentstyle[preprint,aps,prb]{revtex}   

\begin{document}
\tightenlines

\title{Microwave-induced constant voltage steps in
surface junctions of  Bi$_2$Sr$_2$CaCu$_2$O$_{8+\delta}$ single
crystals}

\author{Yong-Joo Doh,$^1$ Jinhee Kim,$^2$ Kyu-Tae Kim,$^2$ and Hu-Jong Lee$^1$}
\address{$^1$Department of Physics,
Pohang University of Science and Technology,\\ Pohang, 790-784,
Korea}
\address{$^2$Electricity Group, Korea Research Institute of Standards
and Science,\\ Yusong-gu, Taejon, 305-600, Korea}

\maketitle
\
\

\begin{abstract}
We have observed the zero-crossing steps in a surface junction of
a mesa structure micro-fabricated on the surface of a
Bi$_2$Sr$_2$CaCu$_2$O$_{8+\delta}$ single crystal. With the
application of microwave of frequencies 76 and 94 GHz, the
current-voltage characteristics show clear voltage steps
satisfying the ac Josephson relation. Increasing the microwave
power, the heights of the steps show the Bessel-function behavior
up to step number $n=4$. We confirm that the intrinsic surface
junction meets the criterion for the observation of zero-crossing
steps.
\end{abstract}

Irradiated with a microwave of frequency $f$, Josephson tunnel
junctions can spontaneously exhibit quantized dc voltages of
$V_n=nhf/2e$ in the absence of a bias current, where $n$ is an
integer and $h$ is the Planck's constant.\cite{LangenChen} In
current-voltage ($I$-$V$) characteristics, this effect manifests
itself as constant voltage steps crossing the zero-current
axis.\cite{Shapiro} The occurrence of these voltage steps is a
direct consequence of the ac Josephson effect and the
phase-coherent pair tunneling in response to an external
electromagnetic excitation.\cite{Josephson} Since no voltages
other than the quantized values $V_n$ are present for zero current
bias, Josephson tunnel junctions are ideal as voltage standards
which require constant voltage output independent of environmental
parameters such as temperature or humidity.\cite{Levin77} Thus,
most Josephson voltage standards currently in use consist of
several thousands of Nb/AlO$_x$/Nb tunnel junctions connected in
series, with each junction exhibiting highly hysteretic $I$-$V$
characteristics.\cite{Niem84,Popel}

The highly anisotropic high-$T_c$ superconductors (HTSCs) with
layered structures, such as Bi$_2$Sr$_2$CaCu$_2$O$_{8+\delta}$
(Bi-2212) and Tl$_2$Ba$_2$Ca$_2$Cu$_3$O$_{10+\delta}$ (Tl-2223),
can be considered as series arrays of Josephson tunnel junctions
along the $c$-axis.\cite{Kleiner} For Bi-2212 single crystals the
superconducting order parameter tends to be localized in $\sim
3$-\AA-thick Cu-O bilayers and the transport along the $c$-axis
occurs mainly via Josephson tunneling between the neighboring Cu-O
bilayers, which are $\sim 12$ \AA~ apart from each other. The
$c$-axis dc $I$-$V$ characteristics of a Bi-2212 single crystal
usually show multiple quasiparticle branches with very large
hysteresis, where the number of branches corresponds to the number
of intrinsic Josephson junctions (IJJs) in the
crystal.\cite{Kleiner,Tanabe,Yurgens,Doh99,Nam99} In spite of the
general consensus for dc $I$-$V$ characteristics, the microwave
response of a stack of IJJs in HTSCs is unclear. Applied with a
microwave, a stack of IJJs has been observed to exhibit constant
voltage steps. However, their values do not satisfy the Josephson
frequency-voltage relation and strongly depend on the microwave
power.\cite{IriePruss} Thus, the authors of Ref.
\onlinecite{IriePruss} have attributed their results to the
microwave-induced phase-locked fluxon motion in a series stack of
IJJs, rather than the ac Josephson effect. Although Shapiro steps
or zero-crossing steps are observed in the studies, the measured
interval between the voltage steps, $\Delta V$, shows a large
difference from the expected value $Nhf/2e$ and depends
sensitively on the frequency and power of the applied
microwave.\cite{Kleiner,John97} Here, $N$ is the total number of
IJJs in a measuring stack. This discrepancy is ascribed to other
coupling mechanisms which give rise to the phase locking of IJJs
in addition to the ac Josephson effect.\cite{KleinSakai}

In this paper, we report an observation of clear zero-crossing
voltage steps in $I$-$V$ characteristics of stacks of IJJs in
Bi-2212 single crystals, irradiated with a microwave of frequency
$f=76$ and $94$ GHz. The voltage steps are believed to be from a
single intrinsic Josephson junction formed on the top surface of
the stack in contact with a metallic (Au) electrode. The voltage
difference between the successive steps coincides with the
expected value of $hf/2e$. The magnitude of the voltage steps
follows the Bessel-function dependence on the applied microwave
power up to the order of $n=4$. This implies that the observed
voltage steps are genuine zero-crossing voltage steps. The
critical current of the surface intrinsic Josephson junction is
significantly suppressed due to the proximity contact to the
normal-metal electrode. This severe reduction of the critical
current allows one to isolate the microwave response of the
``surface junction'' from that of the rest of the ``inner
junctions'' in a stack of Bi-2212 single crystals. From our
experimental results we were able to confirm the necessary
condition for observing zero-crossing steps in an intrinsic
Josephson junction of highly anisotropic high-$T_c$
superconductors.

Stacks of IJJs were fabricated on the surface of a Bi-2212 single
crystal using photolithography and Ar-ion
etching.\cite{Doh99,Nam99}  The Bi-2212 single crystal, grown by
standard solid-state-reaction method,\cite{Chung97} was glued onto
a MgO substrate using negative photoresist and was cleaved using
pieces of Scotch tape until an optically smooth surface appeared.
A thin ($\sim$ 50 nm) Au film was evaporated on top of the crystal
immediately after cleaving to protect the crystal surface from any
contamination during further fabrication processes. Then a large
base mesa ($450\times 20 \times 1.2$ $\mu$m$^3$) was formed by
photolithography and ion-beam etching, using the beam voltage
$V_{\rm beam} = 300$ V and the beam current $I_{\rm beam} = 0.8$
mA/cm$^2$. The ash of photoresist was removed by oxygen plasma
etching. To prevent the regions of the specimen other than the top
surface of the base mesa from being shorted to the contact
electrodes, an insulation layer of photoresist was placed around
the base mesa. Then a 400-nm-thick Au film was further evaporated
and patterned afterwards by photolithography and ion-beam etching
to form electrical extension pads and small stacks ($18 \times 20
\times 0.04$ $\mu$m$^3$) on top of the base mesa. The lateral
dimensions of a small stack were determined by the narrow width of
the base mesa and the breadth of the electrode, which also acted
as a mask for the fabrication of the small stack. The thickness of
the measuring stack, which corresponded to the number of junctions
in it, was controlled by adjusting the ion-beam etching time. The
fabrication procedure was completed by removing the remnant
photoresist by oxygen plasma etching. The heat treatment of the
specimen was limited to $T<$ 120 $^{\rm o}$C during the entire
microfabrication process.

The microwave response of the specimen was measured at $T=4.2$ K.
The microwave generated by a Gunn diode was transmitted through a
waveguide and coupled inductively to the specimen placed at
$\lambda$/4 distance from the end of the waveguide. The maximum
available microwave power was 100 mW for $f=76$ GHz and 50 mW for
$f=94$ GHz. The power coupled to the specimen was tuned by using a
level set attenuator.

Transport measurements were carried out using a three-terminal
measurement method (see the inset in Fig. 1). Shown in Fig. 1 is a
typical $c$-axis resistance vs temperature, $R_c(T)$. The
resistance shows a weak semiconducting behavior above $T_c \approx
87$ K, indicating that the crystal is in a slightly overdoped
regime. One also notices that the resistance remains finite below
$T_c$ with a secondary peak appearing far below $T_c$. It is
attributed to a weak intrisic Josephson junction formed at the
surface of a measuring stack in contact with a Au normal-metal
electrode.\cite{Doh99,Nam99} The superconductivity of the topmost
Cu-O bilayer of a stack is suppressed by the proximity contact to
a normal metal (Au) pad rather than by a degradation effect of the
surface layer.\cite{Doh99} Thus, the surface Cu-O bilayer has a
superconducting transition temperature $T_c^{\prime} \approx 31$ K
far below the bulk $T_c$. In the temperature range of
$T_c^{\prime}<T<T_c$, the surface junction can be considered as a
normal metal/insulator/$d_{x^2-y^2}$-wave-superconductor (NID)
junction\cite{Won94} consisting of the surface Cu-O bilayer in the
normal state and the adjacent inner bilayer in the superconducting
state. Thus, $R_c(T)$ corresponds to a quasiparticle tunneling
resistance of the NID junction with a junction
resistance\cite{Doh99} $R_n^{\prime}=R_c(T_c)=3.9$ $\Omega$. As
the surface junction becomes Josephson coupled below
$T_c^{\prime}$, $R_c(T)$, which is essentially the contact
resistance between Au pad and the topmost Cu-O bilayer, becomes
less than 40 m$\Omega$ in our specimen.\cite{Annealing}

Figure 2 shows the $I$-$V$ characteristics of IJJs in a stack
below $T_c^{\prime}$ in the absence of an exteranl rf field. With
increasing bias current just above the critical current of each
intrinsic Josephson junction in a stack, periodic voltage jumps
occur in units of $V_c \approx 23$ mV, and the $I$-$V$ curves show
highly hysteretic behavior. Although not apparent in the figure,
the number of quasiparticle branches in the $I$-$V$
characteristics indicates that 28 IJJs are contained in the
measuring stack. The average critical current $I_c$ is about $4.5$
mA and the normal state junction resistance of the inner
junctions, estimated from the linear portion of the $I$-$V$
curves, is $R_n=0.7$ $\Omega$. The inset of Fig. 2 shows the
enlarged view of the $I$-$V$ curves in the low bias region. One
notices that the weak surface junction shows a much smaller
critical current $I_c^{\prime} \approx 130$ $\mu$A with clear
hysteresis. The reduced critical current of the surface junction,
compared to the ones of the inner junctions, is due to the
suppressed superconductivity of the surface layer. This result is
consistent with the finite-resistance behavior of the $R_c(T)$
curve in Fig. 1.\cite{Doh99,Nam99}

Figure 3(a) shows the $I$-$V$ characteristics of the specimen with
the application of a microwave of frequency $f=94$ GHz. Clearly
seen are the two steps of height $\Delta I_1 = 120$ $\mu$A. The
voltage difference between the two steps is about $400 \pm 20$
$\mu$V in agreement with the expected value of $\Delta
V=2hf/2e=389$ $\mu$V, implying that these steps are genuine
zero-crossing steps corresponding to the step number $n=\pm 1$.
Due to the weakness of the transmitted microwave power, we could
not observe other steps of $n > 3$ for $f=94$ GHz.

Shown in Fig. 3(b) are the $I$-$V$ characteristics of the same
stack at a microwave frequency $f=76$ GHz. Compared to the case of
$f=94$ GHz, the height of the $n=1$ step becomes reduced while the
steps of higher orders $n= \pm 2, \pm 3$ are seen more clearly.
The steps of $n \geq 3$ do not cross the zero-current line,
possibly due to a large leakage current. The remarkably large
leakage current for the $c$-axis tunneling in Bi-2212 high-$T_c$
superconductors is attributed to the existence of the gapless node
for the $d_{x^2-y^2}$-wave order parameter. By increasing the
microwave power, we were able to identify other voltage steps up
to the order of $n=4$. Further increase of the microwave power
caused a noticeable slope in the voltage steps, possibly due to
chaotic switching of the surface junction between a
Josephson-tunneling state and a resistive one.\cite{KM85} Once any
of the Josephson junctions in the stack becomes resistive, all the
steps are bound to exhibit a finite resistive slope, making it
difficult to identify the steps at high bias voltages.

Figure 4 shows the measured step heights as a function of the
square root of the applied microwave power, $P^{1/2}$. Varying the
step order from $n=0$ to 4, the measured step heights are in
qualitative agreement with the relation of $\Delta I_n =
I_c^{\prime}|J_n(I_{ac}f_c^{\prime}/I_c^{\prime}f)|$, where
$I_c^{\prime}$ is the critical current of the surface junction at
$T=4.2$ K, $J_n$ the $n$-th order Bessel function, $I_{ac}$ the
applied rf current, and
$f_c^{\prime}=2eI_c^{\prime}R_n^{\prime}/h$ the characteristic
frequency of the surface junction.\cite{Barone} We obtained the
fitting parameter $I_c^{\prime}$ (4.2 K) to be $180$ $\mu$A, which
corresponds to the critical current density of 50 A/cm$^2$. This
value is consistent with the ones observed in the surface
junctions of other specimens at 4.2 K.\cite{Doh99,Nam99} To reveal
the Bessel-function behavior, a Josephson junction is required to
satisfy the condition\cite{KM85} of $\Omega^{2}\beta = (f/f_p)^2
\gg 1$, where $\Omega=f/f_c$ is the frequency reduced with the
characteristic frequency of a junction, $\beta=2eI_cR_n^2C/\hbar$
the hysteresis parameter,\cite{Barone} and $f_p =\sqrt{eI_c/\pi h
C}$ the Josephson plasma frequency. $C$ is the capacitance of the
Josephson junction. This criterion is not satisfied with the {\it
inner} IJJs in the stack but is satisfied with the {\it surface}
junction (see Table I). Thus we infer that the observed
zero-crossing steps should originate from the weak surface
junction. As shown in Table I, specimens used by other groups
concerning the microwave-induced fluxon motion\cite{IriePruss} or
collective behavior of the IJJs\cite{Kleiner,John97} do not meet
the above criterion. Nonetheless, one can notice that the typical
parameters of Josephson junctions currently used for Nb-based
voltage standards\cite{Popel} are similar to the ones of the
surface junction. Although the observing conditon\cite{KM85} for
the zero-crossing steps was originally proposed for Josephson
tunnel junctions made of conventional superconductors, our results
indicate that the IJJs in HTSCs provide high potential for
observing the same phenomenon.

The observing condition for the zero-crossing steps can be
rewritten as $\Omega^2\beta = (\pi h/e)(\epsilon f^2/dJ_c) \gg 1$,
where $\epsilon$ is a dielectric constant of the blocking layer
between adjacent conducting bilayers, $d$ a inter-bilayer
distance, and $J_c$ a critical current density of the intrinsic
Josephson junction. For our specimen in this study, the critical
current density of the inner junctions is 24 times larger than
that of the surface junction. An inner junction thus has the
Josephson plasma frequency $f_p$ about five times larger than that
of the surface junction. IJJs with larger critical current
densities require higher microwave frequencies and higher power to
produce stable zero-crossing steps. Reducing the tunneling
critical current density is, therefore, required to obtain the
stable voltage steps from the inner IJJs in a stack. In addition,
to prohibit any nonuniform rf-current flow into the junction,
which becomes more probable at higher microwave frequencies, one
needs to reduce the junction size. These requirements may be
fulfilled with ultra-small IJJs in Bi-2212 single
crystals\cite{Latyshev} or with IJJs in Bi-2212 single crystals
intercalated with guest molecules such as HgI$_2$ or
HgBr$_2$.\cite{HgBi}

In summary, we have studied the inverse ac Josephson effect from a
intrinsic Josephson junction located in the surface of Bi-2212
single crystals irradiated with the external microwave of $f=76$
and $94$ GHz. The surface weak Josephson junction shows clear
voltage steps satisfying the ac Josephson relation and the step
heights follow the Bessel-function behavior with increasing
microwave power, up to the step number $n=4$. Our results indicate
that the intrinsic Josephson junctions in highly anisotropic HTSCs
with very low tunneling critical current density may be a
promising candidate for the observation of zero-crossing steps.

This work was supported in part by KRISS Project No. 98-0502-102.
This work was also supported by BSRI under Contract No.
1NH9825602, MARC under Contract No. 1MC9801301, and POSTECH under
Contract No. 1UD9900801.

\begin{table}[hbt]
\caption{The junction parameters of the {\it surface} junction and
the {\it inner} junctions for our specimen and those of Bi-2212
single crystals and Nb/Al$_2$O$_3$/Nb used in other works. $J_c$
is the critical current density of the junction, $f_p$ the
Josephson plasma frequency, $\Omega$ the reduced frequency, and
$\beta$ the hysteresis parameter or the Stewart-McCumber
parameter.}

\label{table1}
\begin{tabular}{|c|ccccc|}
   & $J_c$ (A/cm$^2$) & $f_p$ (GHz)  & $f$ (GHz) & $f/f_p$ & $\Omega^2\beta$ \\ \hline
  {\it Surface} Junction & 50 & 22 & 76 & 3.5 & 12.3 \\
  {\it Inner} Junction & 1200 & 110 & 76 & 0.69 & 0.48 \\ \hline
  As-grown\cite{IriePruss} & 2200 & 150 & 12 & 0.08 & 0.006 \\
  Ar-annealed\cite{Kleiner} & 150 & 40 & 3.1 & 0.078 & 0.006 \\
  O$_2$-annealed\cite{Kleiner} & 1050 & 100 & 3.2 & 0.032 & 0.001 \\
  Air-annealed\cite{John97}& 650 & 80 & 31 & 0.39 & 0.15 \\ \hline
  Nb/Al$_2$O$_3$/Nb\cite{Popel} & 33 & 21 & 70 & 3.3 & 10.9 \\
\end{tabular}
\end{table}

\begin{figure}
 \caption{The $c$-axis resistance $R_c(T)$ which was obtained using a
 three-terminal measurement method. The bulk superconducting transition
 temperature $T_c$ is $\sim 87$ K and the suppressed superconducting
 transition temperature of the surface layer $T_c^{\prime}$ is $\sim 31$ K.
 The contact resistance is not subtracted. Inset: a schematic
 configuration of the measurements.}
\end{figure}

\begin{figure}
 \caption{The $I$-$V$ characteristics of the specimen at $T=15$
 K without external microwave irradiation. The critical currents of the inner
 junctions are $I_c = 3.5 \sim 5.0$ mA. Inset: magnified view of the
 low bias region, showing clear hysteresis in the $I$-$V$ characteristics
 of the surface junction. The critical current of the surface
 junction $I_c^{\prime}$ is about 130 $\mu$A.}
\end{figure}

\begin{figure}
 \caption{The $I$-$V$ characteristics showing zero-crossing voltage steps
 with application of a microwave of (a) $f=94$ GHz and (b) $f=76$
 GHz at $T=4.2$ K. Each division in the vertical axis is 40 $\mu$A
 and in the horizontal axis is 1 mV.}
\end{figure}

\begin{figure}
 \caption{Measured step heights, from the step number $n=0$ to $4$, as
 a function of the square root of the microwave power for $f=76$ GHz.
 The solid lines are fits to
 $\Delta I_{n}=I_{c}^{\prime}|J_n(I_{ac}f_{c}^{\prime}/I_{c}^{\prime}f)|$
 with $I_c^{\prime}$ (4.2 K) $= 180$ $\mu$A as a fitting parameter.}
\end{figure}


\begin{references}
\bibitem{LangenChen} D. N. Langenberg {\it et al.}, Phys. Lett.
{\bf 20}, 563 (1966); J. T. Chen, R. J. Todd, and Y. W. Kim, Phys.
Rev. B {\bf 5}, 1843 (1972).
\bibitem{Shapiro} S. Shapiro, Phys. Rev. Lett. {\bf 11}, 80
(1963); S. Shapiro, A. R. Janus, and S. Holly, Rev. Modern. Phys.
{\bf 36}, 223 (1964).
\bibitem{Josephson} B. D. Josephson, Phys. Lett. {\bf 1}, 251
(1962); Rev. Modern. Phys. {\bf 36}, 216 (1964).
\bibitem{Levin77} M. T. Levinsen {\it et al.}, Appl. Phys. Lett. {\bf 31},
776 (1977).
\bibitem{Niem84} J. Niemeyer, J. H. Hinken, and R. L. Kautz, Appl.
Phys. Lett. {\bf 45}, 478 (1984).
\bibitem{Popel} R. P\"{o}pel, Metrologia {\bf 29}, 153 (1992) and
references therein.
\bibitem{Kleiner} R. Kleiner {\it et al.}, Phys. Rev. Lett. {\bf 68},
2394 (1992); R. Kleiner and P. M\"{u}ller, Phys. Rev. B {\bf 49},
1327 (1994).
\bibitem{Tanabe} K. Tanabe {\it et al.}, Phys. Rev. B {\bf 53}, 9348
(1996); M. Itoh, S. Karimoto, K. Namekawa, and M. Suzuki, $ibid.$
{\bf 55}, R12001 (1997).
\bibitem{Yurgens} A. Yurgens {\it et al.}, Phys. Rev. B {\bf 53}, R8887
(1996); Phys. Rev. Lett. {\bf 79}, 5122 (1997).
\bibitem{Doh99} Y.-J. Doh and H.-J. Lee, to appear in
{\it Proceedings of the 22nd International Conference on Low
Temperature Physics, Helsinki, Finland, 1999}; Y.-J. Doh, H.-J.
Lee, and H.-S. Chang, cond-mat/9907251.
\bibitem{Nam99} N. Kim, Y.-J. Doh, H.-S. Chang, and H.-J. Lee,
Phys. Rev. B {\bf 59}, 14639 (1999).
\bibitem{IriePruss} A. Irie and G. Oya, Physica C {\bf 293}, 249
(1997); W. Prusseit {\it et al.}, Physica C {\bf 293}, 25 (1997).
\bibitem{John97} H. L. Johnson {\it et al.}, J. Appl. Phys. {\bf
82}, 756 (1997).
\bibitem{KleinSakai} R. Kleiner {\it et al.}, Phys. Rev. B {\bf 50},
3942 (1994).
\bibitem{Chung97} M. Chung {\it et al.}, J. Kor. Phys. Soc. {\bf 31},
384 (1997).
\bibitem{Won94} H. Won and K. Maki, Phys. Rev. B {\bf 49}, 1397
(1994).
\bibitem{Annealing} No surface junction was observed in the work of
Ref. \onlinecite{Tanabe} in which the specimens were annealed
after Au deposition. This may be due to material inter-diffusion
between the Au film and the surface layer of Bi-2212 single
crystals.
\bibitem{Barone} A. Barone and G. Paterno, {\it Physics and
Applications of the Josephson Effect} (Wiley, New York, 1982).
\bibitem{KM85}R. L. Kautz, Appl. Phys. Lett. {\bf 36}, 386
(1980); R. L. Kautz and R. Monaco, J. Appl. Phys. {\bf 57}, 875
(1985).
\bibitem{Latyshev} Yu. I. Latyshev {\it et al.}, Phys. Rev. Lett.
{\bf 82}, 5345 (1999).
\bibitem{HgBi} M. Lee, H.-S. Chang, Y.-J. Doh, H.-J. Lee, W. Lee,
J.-H. Choy, and D. H. Ha, to appear in Physica B; A. Yurgens {\it
et al.}, cond-mat/9907159.

\end{references}
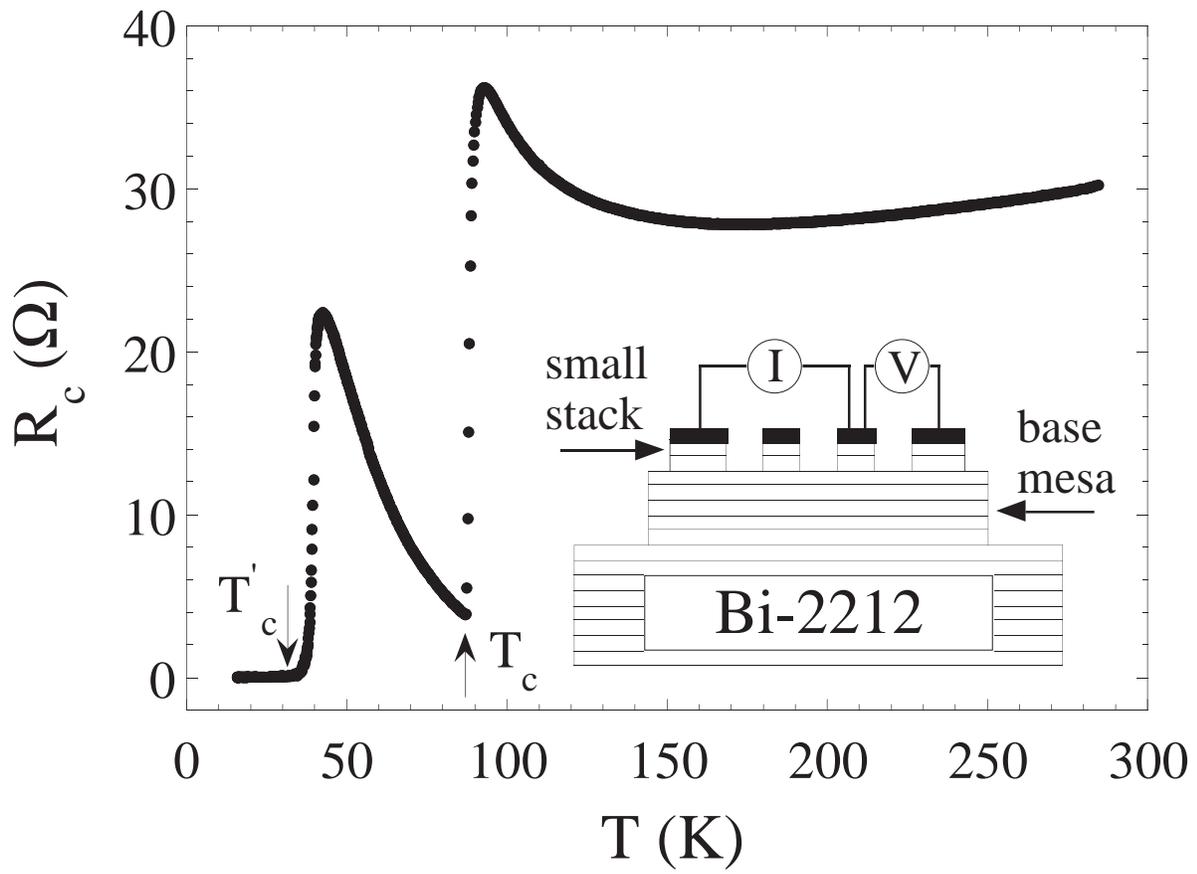
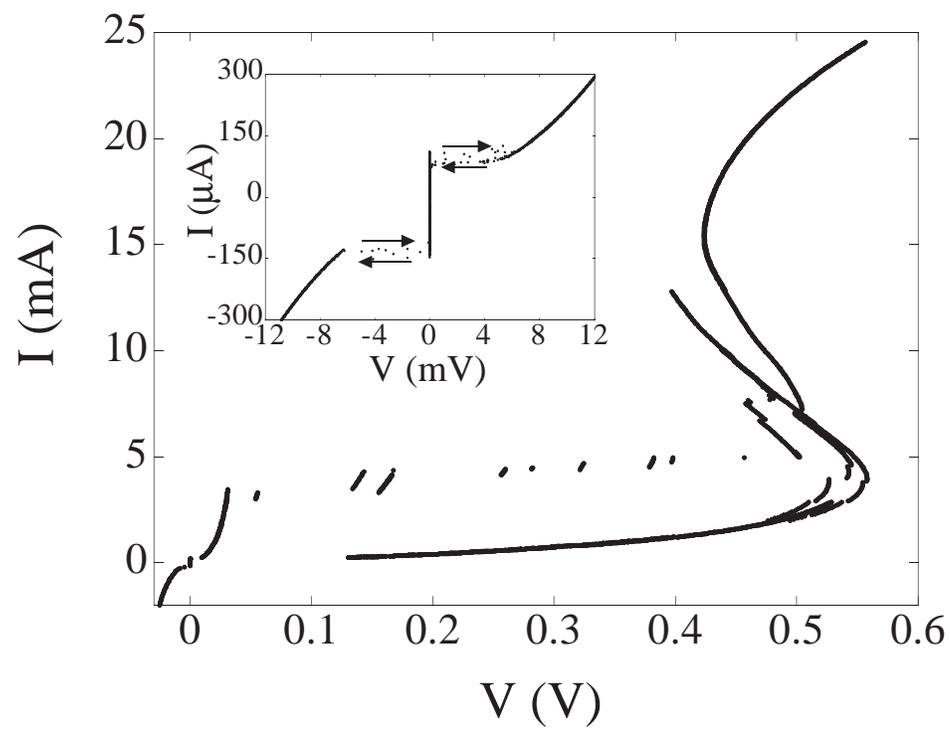
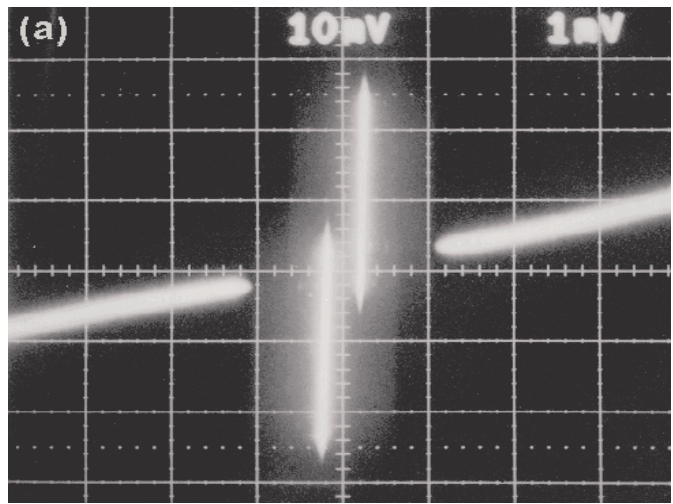
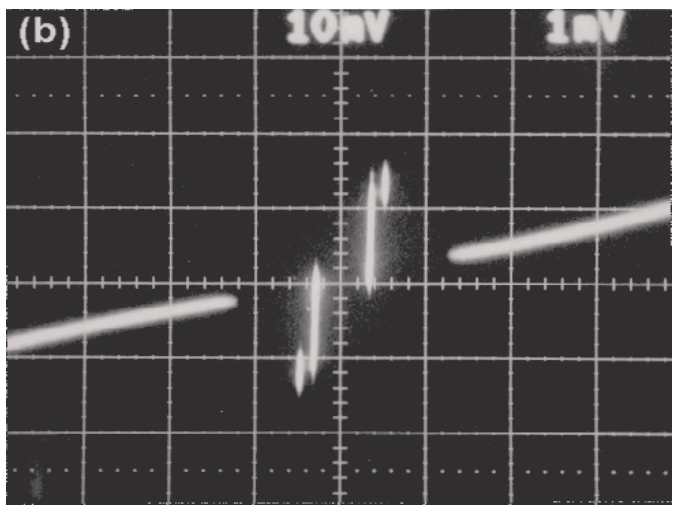
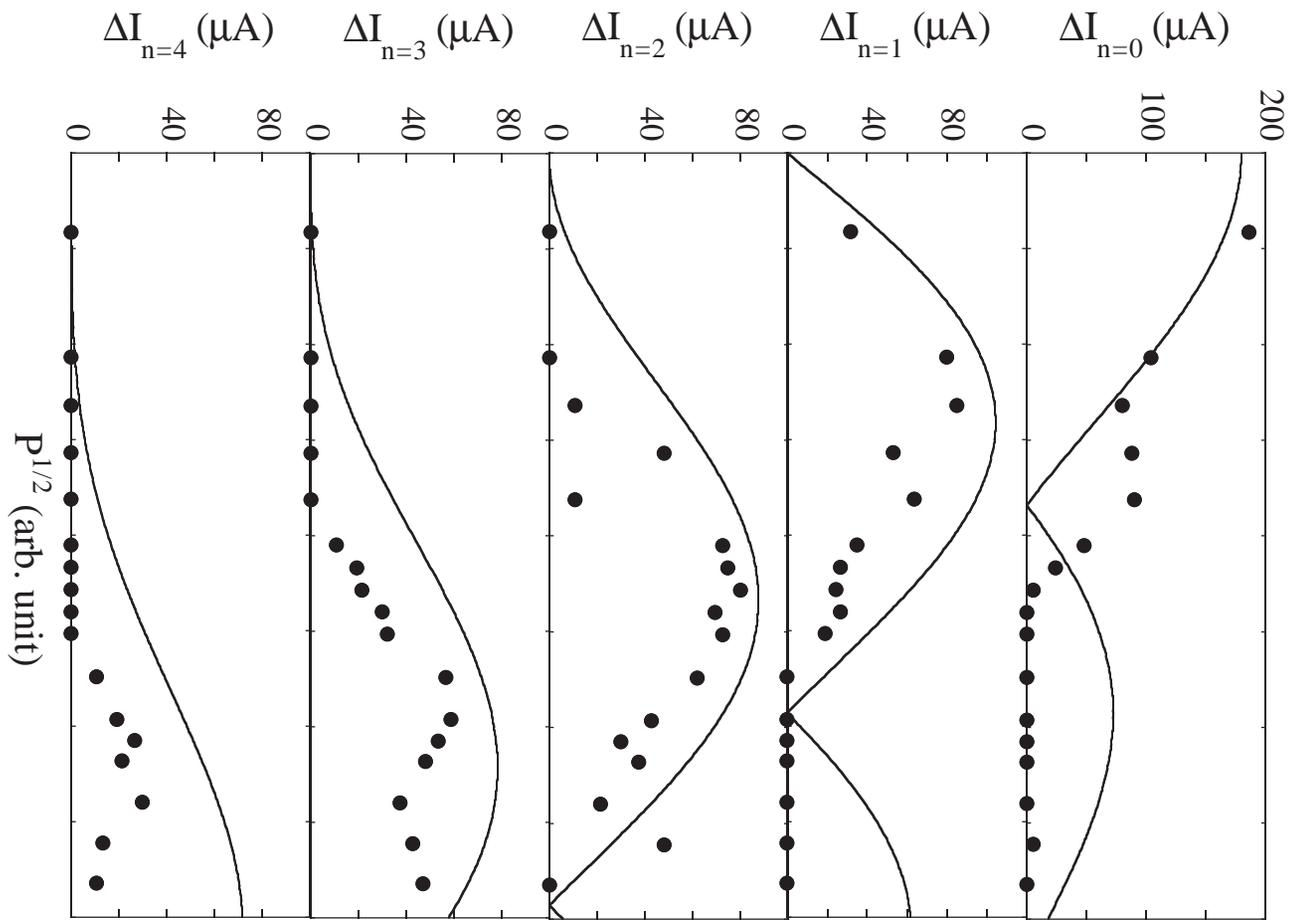
\end{document}